\documentclass[12pt,a4paper,onecolumn]{article}

\usepackage[T1]{fontenc}
\usepackage[utf8]{inputenc}
\usepackage{graphicx}
\usepackage{authblk}
\usepackage{xcolor}

\title{Plasmon-polariton modes in fullerenes.}
\author[1,2]{N.L. Matsko}

\affil[1]{\footnotesize Moscow Institute of Physics and Technology, Dolgoprudny, Moscow Region, 141701, Russia}
\affil[2]{P.N. Lebedev Physical Institute, Russian Academy of Sciences, Leninskii prosp. 53, 119991 Moscow, Russia}

\date{}

\begin{document}

\maketitle

\begin{abstract}
It is well known that collective electronic excitations in fullerene C$_{60}$ are manifested as Mie plasmons, and in graphene (the limiting case of an infinitely large fullerene) collective excitations are of plasmon-polariton type. How the properties of plasmons change in fullerenes with intermediate sizes is poorly understood. This problem is considered in current paper in the framework of the GW approximation on the example of fullerenes C$_{60}$, C$_{240}$ and C$_{540}$. The calculations show that a high-frequency plasmon-polariton resonance begins to form in C$_{240}$, and in C$_{540}$ the intensity of this resonance becomes comparable to the intensity of Mie plasmon resonance. Thus with a further increase in fullerene size, a gradual transition from Mie plasmons to plasmon-polaritons should be observed.
\end{abstract}

\section{Introduction}

Fullerenes are promising objects for use in nanoplasmonics. Despite the geometric simplicity of the molecule, collective electronic excitations in fullerenes have a non-trivial structure. That is due to the excitation of plasmons in both $\pi$ and $\sigma$ electron orbitals, the interaction of these plasmons with each other and with single-electron excitations \cite{sohmen,ju,verkh1}.
Experiments on photoabsorption and electron energy loss spectroscopy in C$_{60}$ show two dominant peaks in the region of 6 and 20 eV, which correspond to the localized surface plasmon resonance (LSPR) of $\pi$ and $\sigma$ electrons respectively \cite{sohmen, saito,hertel, kafle}. These resonances are dipole, optically active oscillations of the electron density. In the framework of the Mie theory, LSPR corresponds to excitations with angular momentum $l=1$ \cite{mie}. Besides, in a number of papers the plasma oscillations in C$_{60}$ with $l>1$ were considered \cite{ju,gerchikov,verkh2}. However, the significance of such multipole plasmons in the excitation spectrum of the studied fullerenes is controversial, thus we will not consider them in this paper.

With an increase in the size of the fullerene molecule, the picture of plasma oscillations should tend to that in graphene. Plasma excitations in graphene are manifested as surface plasmon-polaritons (SPPs) which are traveling electron density waves \cite{politano,cui,gei-nov}. The two characteristic plasmon resonances with frequencies of 4.7 and 14.6 eV are observed in graphene. They correspond to the SPP oscillations of $\pi$ and $\sigma$ electrons respectively.

It is worth to note some differences between LSPR and SPP plasmons. LSPR dipole mode is the translational excitation, where the electron density oscillates relative to the center of mass of the ion system. SPP is the compressional excitation of electron density wave propagating along the structure's surface (fig. 1). Unlike Mie plasmons, SPP modes can't be excited by a spherically symmetric perturbation. Therefore SPPs are not observed in photoabsorption experiments, but can be excited by passing charged particles in the electron energy loss spectroscopy measurements \cite{verkh2,connerade}. 

\begin{figure}[h]
\centering
\includegraphics[width=0.5\textwidth]{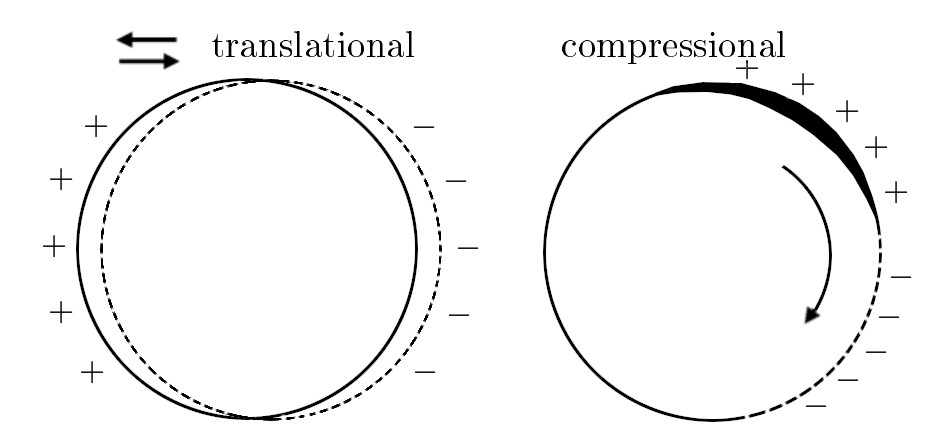}
\caption{Translational plasmons (LSPR) and compressional plasmons (SPP) in fullerene structure.}
\end{figure}

To summarize, the transition from plasma excitations of the LSPR type to excitations of the SPP type with an increase in the fullerene size is of great interest. In addition, information on plasma excitations in fullerenes larger than C$_{60}$ is rather scarce \cite{ruiz,choi,verkh3}. The purpose of our work is to fill this gap. Below we will consider how the spectrum of plasma excitations changes within the GW approximation for C$_{60}$, C$_{240}$ and C$_{540}$ fullerenes. In particular, the formation of SPP oscillations will be studied. A similar study on the formation of SPP mode in sodium nanoparticles was done in our work \cite{napla}.

\section{Computational details}

DFT calculations in the Quantum Espresso code \cite{qe} with PBE GGA pseudopotential were used as a starting point for the one-iteration G$_0$W$_0$. Because of high computational requirements, the calculation precision for C$_{540}$ was lower than for C$_{60}$ and C$_{240}$. The plane wave basis cutoff energy was set to 45 Ry for C$_{60}$ and C$_{240}$ and to 30 Ry for C$_{540}$. Computations were performed for cubic supercell geometry with the side length of 60 Bohr.
For GWA calculations of the system response function the BerkeleyGW \cite{hyb-Lou},\cite{deslippe},\cite{bgw2} package was applied. The energy range for unoccupied bands was set to 15 eV for C$_{60}$ and C$_{240}$ and to 13 eV for C$_{540}$ (2400 unoccupied bands for C$_{540}$). Full frequency dependence method with contour-deformation formalism for the inverse dielectric matrix calculations was used. Energy cutoff for the dielectric matrix was set to 4.0 Ry for C$_{60}$ and C$_{240}$ and to 3.0 Ry for C$_{540}$. The Coulomb interaction was cutoff on the edges of cell box.
The given parameters provide frequency convergence with an accuracy approximately of 0.1 eV for C$_{60}$ and C$_{240}$ and 0.5 eV for C$_{540}$.

To calculate plasmons in the studied objects, the loss function $Im[\epsilon^{-1}(r,r',\omega)]$ was constructed.
The function $\epsilon^{-1}(r,r',\omega)=1+\rho_{ind}(r,\omega)/\rho_{ext}(r')$ describes the response of the system at $r$ to a perturbation at $r'$ at the frequency $\omega$ ($\rho_{ind}$ and $\rho_{ext}$ are induced density and external density perturbation).
For $\epsilon=\epsilon_1+i\epsilon_2$ we can write $Im[\epsilon^{-1}]=-\epsilon_2/(\epsilon_1^2+\epsilon_2^2)$. The denominator $(\epsilon_1^2+\epsilon_2^2)$ is responsible for function poles. Thus, collective electronic excitations (where $\epsilon_1^2+\epsilon_2^2\to0$) are manifested as the corresponding peaks of the loss function \cite{pines},\cite{sturm},\cite{haque},\cite{hanke}. The inverse dielectric function $\epsilon^{-1}_{GG'}(q,\omega)$ was calculated in the G$_0$W$_0$ approximation.  After that the function $\epsilon^{-1}(r,r',\omega)$ was constructed \cite{napla}:

\begin{equation} \label{f_1} \epsilon^{-1}(r,r',\omega)=\sum_{q,G,G'} e^{i(q+G)r}\epsilon^{-1}_{GG'}(q,\omega)e^{-i(q+G')r'} \end{equation}

Calculations were carried out at the gamma point $q=0$.

Furthermore, we can construct $\epsilon^{-1}(r,r',\tau)$ for a certain frequency interval using the windowed Fourier transform with the window function $W(\omega-\omega')$. Since we are interested in the loss function distribution at  plasmon resonance frequencies, we take the window function as $\delta(\omega-\omega_i)$, where $\omega_i$ - LSPR or SPP resonance frequency:

\begin{equation} \label{f_2} \int\limits_{-\infty}\limits^{\infty} \epsilon^{-1}(\omega) W(\omega-\omega') e^{i\omega \tau} d\omega = \int\limits_{-\infty}\limits^{\infty} \epsilon^{-1}(\omega) \delta(\omega-\omega_i) e^{i\omega \tau} d\omega = \epsilon^{-1}(\omega_i)e^{i\omega_i \tau} \end{equation}

Setting $\tau=0$ we get $\epsilon^{-1}(r,r',\omega_i)$. Thus, for $r$ lying on the structure's surface the $\epsilon^{-1}(r,r',\omega_i)$ shows the distribution of $\rho_{ind}(r)$ at the resonance frequency $\omega_i$ at some "zero" moment of time. $Im[\epsilon^{-1}(r,r',\omega_i)]=Im[\frac{\rho_{ind}(r,\tau=0)}{\rho_{ext}(r')}]$, so the loss function is proportional to  $\rho_{ind}(r,\tau=0)$. Thereby, dependence of the $Im[\epsilon^{-1}(r,r',\omega_i)]$ on $r$ allows to analyze the spatial distribution of the phase of the plasmonic excitations under study.
For example, $\rho_{ind}(r)$ should has different signs at opposite points of the fullerene in case of LSPR dipole mode (fig. 1). In case of SPP oscillations the sign of $\rho_{ind}(r)$ varies depending on the phase of the wave at the point $r$ on the fullerene surface.

\section{Results and discussion}

It is worth to point, that from simple expressions for plasma frequencies in the framework of jellium model, $\omega_{LSPR}=\omega_{pl}/\sqrt{3}$ and $\omega_{SPP}=\omega_{pl}/\sqrt{2}$ (where $\omega_{pl}$ is the bulk plasmon frequency of the system). This means that the ratio of SPP to LSPR frequency corresponds to $\sqrt{3/2}$ ($\approx$1.22). So for the fullerenes it is logical to expect the emergence of the SPP resonance at frequencies approximately $\sqrt{3/2}$ greater than the corresponding LSPR frequency.

\begin{figure}[h]
\centering
\includegraphics[width=0.75\textwidth]{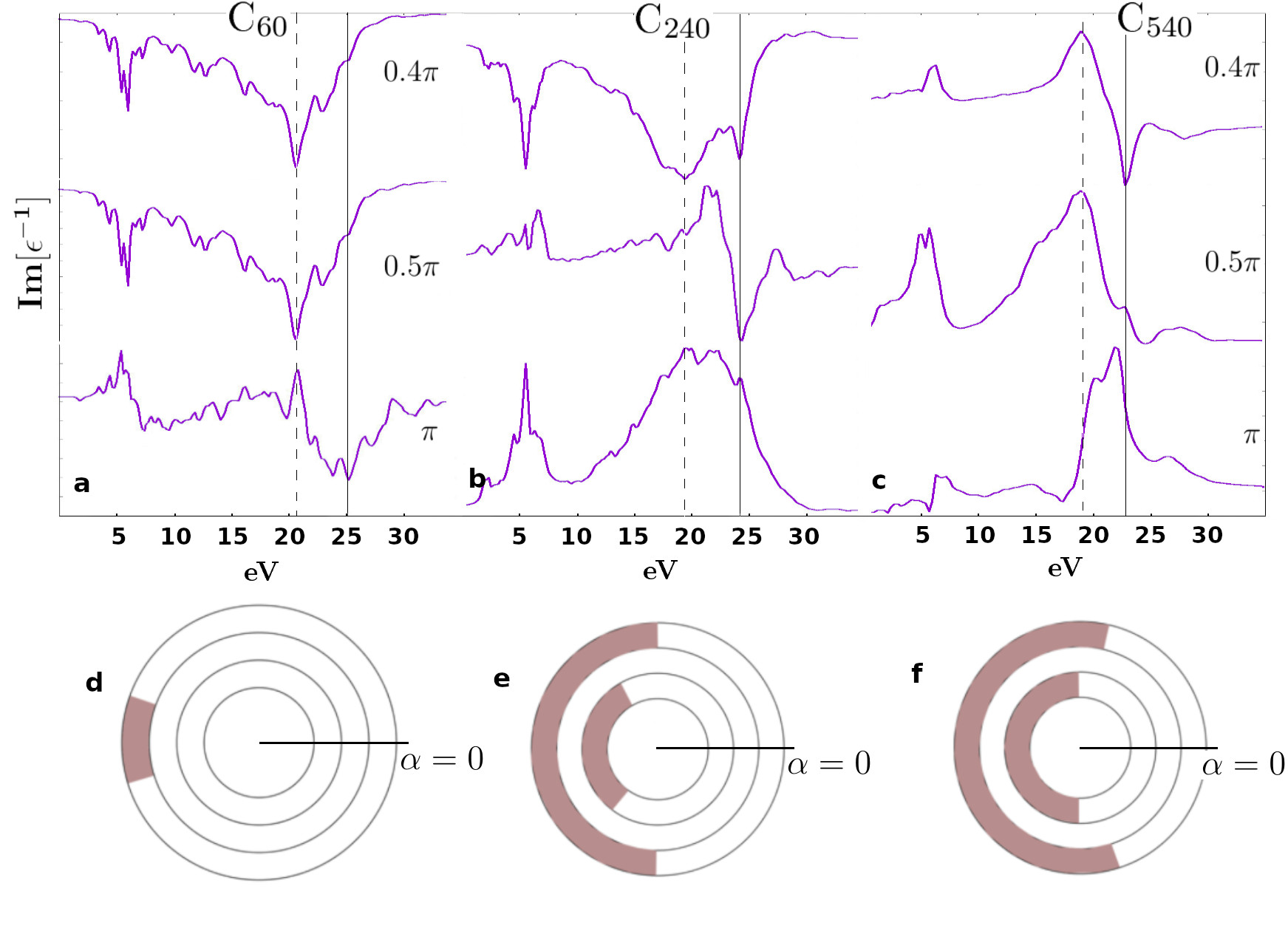}
\caption{a, b, c - $Im[\epsilon^{-1}(r,r',\omega)]$ in C$_{60}$, C$_{240}$, C$_{540}$ for $\alpha$ angles $0.4\pi$, $0.5\pi$ and $\pi$. Dashed and solid lines mark LSPR and SPP peaks for $\sigma$ electrons respectively. d,e,f - distribution of $\rho_{ind}$ along the surface for C$_{60}$, C$_{240}$, C$_{540}$. The light area of the rings corresponds to $\rho_{ind}<0$, the dark area of the rings corresponds to $\rho_{ind}>0$. The outer rings show the distribution for LSPR, the inner rings show the distribution for SPP mode.}
\end{figure}

Figures 2(a,b,c) show the function $Im[\epsilon^{-1}(r,r',\omega)]$ in C$_{60}$, C$_{240}$ and C$_{540}$ for three values of $\alpha$. $\alpha$ is an angle between points $r$ and $r'$ located on the structure's surface with a vertex in the center of structure. The functions are shown for angles $0.4\pi$, $0.5\pi$ and $\pi$, which are chosen for a clear representation of plasmonic spatial phase distribution. Dashed and solid lines mark LSPR and SPP peaks for $\sigma$ electrons respectively.
In C$_{60}$ (Fig. 2a) two dominant resonances at frequencies of 6 and 20.6 eV are clearly visible for all angles $\alpha$. These resonances correspond to LSPR excitations of $\pi$ and $\sigma$ electrons respectively. For $\alpha=\pi$, a new peak at frequency of 25.3 eV is seen. The peak is strongly blurred, weakly noticeable at other angles, and retains its sign for all $\alpha$ values. It's frequency is $\approx$1.23 times greater than $\sigma$ LSPR one. We associate this peak with the incipient SPP excitations of $\sigma$ electrons. In C$_{240}$ (Fig. 2b) frequencies of Mie plasmons are 5.6 and 19.4 eV for $\pi$ and $\sigma$ electrons respectively. The SPP resonance has a frequency of 24.2 eV and becomes clearly distinguishable at any angles $\alpha$.  The ratio $\omega_{SPP}/\omega_{LSPR}\approx1.25$.
All three resonances are shifted in comparison with C$_{60}$ to the region of lower frequencies, which is explained by the size effect \cite{ekardt,xiang,sipla}.
As mentioned above, the results for C$_{540}$ are rather qualitative. The peak of SPP resonance lies in the range of 22-22.6 eV, LSPR frequencies for $\pi$ and $\sigma$ orbitals lie in the intervals of 5.8-6 eV and 18.7-20.4 eV respectively. Thus, the frequencies of plasma excitations in C$_{540}$ undergo a further redshift, intensity of SPP peak increases.

Figures 2(d,e,f) show the dependence of plasmonic peaks orientation of the function $Im[\epsilon^{-1}(r,r',\omega_i)]$ on $\alpha$.
The light area of the rings corresponds to negative peaks, the dark area of the rings corresponds to positive peaks. Therefore, as discussed in the Computational details section, for a positive perturbation at point $r'$ the light area corresponds to $\rho_{ind}<0$ and the dark area corresponds to $\rho_{ind}>0$ at "zero" moment of time. Outer rings on figures 2(d,e,f) show the $\rho_{ind}$ distribution at LSPR frequency for $\sigma $ electrons (for $\pi$ electrons the distribution is the same), inner rings show the $\rho_{ind}$ distribution for $\sigma$ electrons at the SPP mode frequency.
As can be seen, the region of positive $\rho_{ind}$ in the LSPR mode consistently grows as we go from C$_{60}$ to C$_{240}$ and to C$_{540}$. Apparently, this is due to an increase in the amplitude of the electron displacements relative to the ion subsystem (see Fig. 1).
SPP spatial charge distribution in C$_{60}$ is fairly uniform, only in C$_{240}$ and C$_{540}$  charge distribution corresponding to traveling wave begins to form.

The main result is the appearance of SPP resonance of $\sigma$ electrons in C$_{240}$ and C$_{540}$.
In the frequency range of $\pi$ orbitals resonance (5-6 eV) it is not possible to separate the SPP contribution from the others.
The probable explanation is that the LSPR and SPP excitations of $\pi$ electrons have close frequencies and a strong coupling, eventually giving a low-frequency resonance of a mixed type. Thus, only the $\sigma$ resonance at 23-25 eV corresponds to the pure plasmon-polariton mode.
In C$_{60}$ SPP excitations are strongly damped. SPP peak is blurred, corresponding charge distribution is uniform. With an increase in the fullerene size, resonance conditions for the excitation of SPP mode are formed when an integer number of wavelengths fit into the path along the structure's surface. In C$_{60}$ the SPP wavelength is much larger than the great circle (intersection of the diametral plane with the sphere) equal to 22 \AA. In C$_{540}$ the great circle ($\approx$68\AA) becomes equal to two SPP half-waves (Fig. 2f, inner ring) and a clear peak corresponding to the resonance appears in the loss function.

\section{Conclusions}

As a result, the picture of plasma oscillations in fullerenes can be summarized  as follows.
Valence electrons of carbon atoms in fullerenes consist of two subsystems - $\pi$ and $\sigma$ orbitals. Each atom contains one $\pi$ electron with a density distribution perpendicular to the fullerene surface, and 3 $\sigma$ electrons lying in the fullerene plane. From simple expressions in the framework of jellium model both subsystems exhibit collective excitations with frequencies $\omega_i\sim\sqrt{n_i/m_i^{eff}}$, where $n_i$ is a concentration and $m_i^{eff}$ is an effective mass of electrons in the given subsystem.
Collective oscillations in C$_{60}$  are manifested as two dominant resonances of 6 and 20.6 eV, which are Mie plasmons of $\pi$ and $\sigma$ electrons. The calculations show that with an increase in the fullerene size, a new clear resonance appears at an energy of 24.2 eV for C$_{240}$ and in the range 22-22.6 eV for C$_{540}$. The resonance corresponds to excitations of the plasmon-polariton type, which are compressional electron density waves propagating along the structure's surface. This SPP mode is formed when an integer number of wavelengths fit into the great circle of the structure's surface. From simple expressions for plasma frequencies in the framework of jellium model the ratio of the SPP and LSPR frequencies should correspond to $\sqrt{3/2}$ ($\approx$1.22). In C$_{240}$ for $\sigma$ electrons the frequencies of SPP and LSPR are 24.2 and 19.4 eV respectively (the ratio is $\approx1.25$). In C$_{540}$ (the calculation precision allow us to make rather qualitative estimates) the picture as a whole remains the same. As the fullerene size grows from C$_{60}$ to C$_{240}$ and C$_{540}$, the collective vibration frequencies successively undergo a redshift due to the size effect. With a further increase of the fullerene molecule, one should expect that the frequency of the SPP mode for $\sigma$ electrons will gradually decrease down to 14.6 eV (the frequency of $\sigma$ plasmons in graphene). The intensity of the LSPR modes will drop to complete attenuation. In the low-frequency resonance of $\pi$ electrons, only the SPP contribution with a frequency of 4.7 eV will remain.

\section{Acknowledgements}

Calculations were carried out on the Joint Supercomputer Center of the Russian Academy of Sciences (JSCC RAS).

\bibliography{thesis}

\begin{thebibliography}{99}
\bibitem{sohmen} E. Sohmen, J. Fink and W. Krätschmer, EPL {\bf17}, 51 (1992)
\bibitem{ju} N. Ju, A. Bulgac, J.W. Keller, Phys. Rev. B {\bf48}, 9071 (1993)
\bibitem{verkh1}  A.V. Verkhovtsev, A.V. Korol, A.V. Solov'yov, P. Bolognesi, A. Ruocco and L. Avaldi, J.Phys.B: At.Mol.Opt.Phys. {\bf45}, 141002 (2012)
\bibitem{saito} Y. Saito et al, Jpn. J. Appl. Phys. {\bf30}, L1068 (1991)
\bibitem{hertel} I.V. Hertel, H. Steger, J. de Vries, B. Weisser, C. Menzel, B. Kamke, and W. Kamke, Phys. Rev. Lett. {\bf68}, 784 (1992)
\bibitem{kafle} B.P. Kafle, H. Katayanagi, M. Prodhan, H. Yagi, C. Huang and K. Mitsuke, J. Phys. Soc. Jpn. {\bf77}, 014302 (2008)
\bibitem{mie} G. Mie, Ann. Phys. {\bf330}, 377—445 (1908)
\bibitem{gerchikov} L.G. Gerchikov, A.N. Ipatov, A.V. Solov'yov and Walter Greiner, J. Phys. B: At. Mol. Opt. Phys. {\bf31}, 3065 (1998)
\bibitem{verkh2} A. Verkhovtsev, A. Korol and A. Solov’yov, Eur.Phys.J.D {\bf66}, 253 (2012)
\bibitem{politano} A. Politano and G. Chiarello, Nanoscale, {\bf6}, 10927-10940 (2014)
\bibitem{cui} L. Cui, J. Wang, M. Sun, Reviews in Physics, {\bf6}, 100054 (2021)
\bibitem{gei-nov} T. Eberlein, U. Bangert, R.R. Nair, R. Jones, M. Gass, A.L. Bleloch, K.S. Novoselov, A. Geim, and P.R. Briddon, Phys. Rev. B {\bf77}, 233406 (2008)
\bibitem{connerade} J.P. Connerade, A.V. Solov’yov, Phys. Rev. A {\bf66}, 013207 (2002)
\bibitem{ruiz} A. Ruiz, J. Breton, J.M. Gomez Llorente. Chem.Phys.Let. {\bf389} 191–197 (2004)
\bibitem{choi} J. Choi, E. Chang, D.M. Anstine, M. Madjet and H.S. Chakraborty
Phys. Rev. A {\bf95}, 023404 (2017)
\bibitem{verkh3} A. Verkhovtsev, A. Korol and A. Solov’yov, Eur.Phys.J.D {\bf70}, 221 (2016)
\bibitem{napla} N.L. Matsko, Phys. Chem. Chem. Phys. {\bf22}, 13285-13291 (2020)
\bibitem{qe} P. Giannozzi et al. J.Phys.:Condens.Matter {\bf21}, 395502 (2009)
\bibitem{hyb-Lou} M.S. Hybertsen and S.G. Louie, Phys. Rev. B {\bf34}, 5390 (1986)
\bibitem{deslippe} J. Deslippe, G. Samsonidze, D.A. Strubbe, M. Jain, M.L. Cohen, S.G. Louie, Comp. Phys. Comm. {\bf183}, 1269–1289 (2012)
\bibitem{bgw2} M. Rohlfing and S.G. Louie, Phys. Rev. B {\bf62}, 4927 (2000)
\bibitem{pines} D. Pines, "Elementary excitations in solids", Benjamin, New York-Amsterdam (1963)
\bibitem{sturm} K. Sturm, Electron energy loss in simple metals and semiconductors, Advances in Physics, 31:1, 1-64 (1982)
\bibitem{haque} M.S. Haque \& K.L. Kliewer. Plasmon Properties in bcc Potassium and Sodium. Phys. Rev. B, 7{\bf6}, 2416–2430 (1973)
\bibitem{hanke} W. Hanke, Dielectric theory of elementary excitations in crystals, Advances in Physics, 27:2, 287-341 (1978)
\bibitem{ekardt} W. Ekardt, Phys.Rev.B {\bf31}, 6360 (1985)
\bibitem{xiang} H. Xiang, X. Zhang, D. Neuhauser, G. Lu, J. Phys. Chem. Lett. {\bf5}, 1163-1169 (2014)
\bibitem{sipla} N.L. Matsko, Phys. Chem. Chem. Phys. {\bf20}, 24933-24939 (2018)

\end{thebibliography}
\bibliographystyle{gost705}

\end{document}